\documentclass[%
 aip,
 amsmath,amssymb,
 reprint,%
]{revtex4-1}

\usepackage{graphicx}
\usepackage{dcolumn}
\usepackage{bm}
\usepackage[dvipsnames]{xcolor}
\usepackage{soul}
\usepackage[utf8]{inputenc}
\usepackage[T1]{fontenc}
\usepackage{mathptmx}
\usepackage{etoolbox}
\usepackage{upgreek}
\usepackage[normalem]{ulem}
\usepackage{graphicx}   
\usepackage{tikz}       
\usetikzlibrary{calc}

\makeatletter
\newcounter{emailmark}

\newcommand{\emailnthmark}[1]{%
  \ifcase\numexpr#1-1\relax
    *\or \dag\or \ddag\or \S\or \P\or
    $\|$\or **\or \dag\dag\or \ddag\ddag
  \fi
}

\newcommand{\getemailmark}[1]{%
  \edef\@tempemailkey{emailmarkfor\expandafter\@gobble\string#1}%
  \@ifundefined{\@tempemailkey}{%
    \stepcounter{emailmark}%
    \expandafter\xdef\csname\@tempemailkey\endcsname{\theemailmark}%
  }{}%
  \emailnthmark{\csname\@tempemailkey\endcsname}%
}

\let\@orig@email\email

\renewcommand{\email}[1]{%
  \edef\@tempemailkey{emailmarkfor\expandafter\@gobble\string#1}%
  \@ifundefined{\@tempemailkey}{%
    \stepcounter{emailmark}%
    \expandafter\xdef\csname\@tempemailkey\endcsname{\theemailmark}%
  }{}%
  \expandafter\xdef\csname authormark@\@currentauthor\endcsname{%
    \csname\@tempemailkey\endcsname
  }%
  \@orig@email{#1}%
}

\let\@orig@author\author
\renewcommand{\author}[1]{%
  \gdef\@currentauthor{#1}%
  \@orig@author{#1}%
}

\let\@orig@doauthor\doauthor
\def\doauthor#1#2#3{%
  \edef\@tempauthormark{authormark@#1}%
  \@ifundefined{\@tempauthormark}{%
    \@orig@doauthor{#1}{#2}{#3}%
  }{%
    \@orig@doauthor{#1}{#2}{#3\textsuperscript{\emailnthmark{\csname\@tempauthormark\endcsname}}}%
  }%
}

\def\@email#1#2{%
  \endgroup
  \patchcmd{\titleblock@produce}
    {\frontmatter@RRAPformat}
    {\frontmatter@RRAPformat{%
      \produce@RRAP{\getemailmark{#2}#1\href{mailto:#2}{#2}}%
    }\frontmatter@RRAPformat}
    {}{}%
}

\makeatother

\begin{document}

\preprint{AIP/123-QED}

\title{Sub-10 mK ``In-cell'' Magnetic Refrigeration for Cryogen-free Cryostats}

\author{Alexander M. Donald}
\email{adonald@ufl.edu}
\affiliation{ 
Department of Physics and National High Magnetic Field Laboratory High B/T Facility,\\ 
University of Florida, Gainesville, Florida 32611, USA
}%

\author{Nicolas Silva}
\affiliation{ 
Department of Physics and National High Magnetic Field Laboratory High B/T Facility,\\ 
University of Florida, Gainesville, Florida 32611, USA
}%

\author{Christopher J. Ollmann}%
\affiliation{ 
Department of Physics and National High Magnetic Field Laboratory High B/T Facility,\\ 
University of Florida,  Gainesville, Florida 32611, USA
}%

\author{Roch Schanen}
\affiliation{%
Department of Physics, Lancaster University, Lancaster LA1 4YB, United Kingdom
}%

\author{Chao Huan}
\affiliation{ 
Department of Physics and National High Magnetic Field Laboratory High B/T Facility,\\ 
University of Florida,  Gainesville, Florida 32611, USA
}%

\author{Sangyun Lee}
\affiliation{ 
Department of Physics and National High Magnetic Field Laboratory High B/T Facility,\\ 
University of Florida,  Gainesville, Florida 32611, USA
}%

\author{Dominique~Laroche}
\affiliation{ 
Department of Physics and National High Magnetic Field Laboratory High B/T Facility,\\ 
University of Florida,  Gainesville, Florida 32611, USA
}%

\author{Richard P. Haley}
\affiliation{%
Department of Physics, Lancaster University, Lancaster LA1 4YB, United Kingdom
}%

\author{Mark W. Meisel}
\affiliation{ 
Department of Physics and National High Magnetic Field Laboratory High B/T Facility,\\ 
University of Florida,  Gainesville, Florida 32611, USA
}%

\author{Rasul Gazizulin}
\email{r.gazizulin@ufl.edu}
\affiliation{ 
Department of Physics and National High Magnetic Field Laboratory High B/T Facility,\\
University of Florida,  Gainesville, Florida 32611, USA
}%

\date{\today}

\begin{abstract}
A design and implementation of ``in-cell`` magnetic refrigeration to achieve sub-10 mK temperatures 
\textit{T} in cryogen-free dilution refrigerators is presented. The ultra low temperatures below 5 mK 
are attained in finite magnetic fields \textit{B} up to 1 T. The holding time below 5 mK varies between 
about 3 to 30 hours, depending on the final magnetic field after demagnetization process. The developed technique can be used to study 
low dimensional devices at ultra low electron temperatures in the High \textit{B}/\textit{T} regime. 
\end{abstract}

\maketitle

%

\section{\label{sec:Intro}Introduction\protect}

Advances in cryogen-free technologies have greatly expanded the availability of millikelvin temperature platforms 
to a broad range of quantum research. However, cooling electrons in low-dimensional devices, 
such as graphene or GaAs heterostructures, is challenging due to weak electron-phonon coupling 
at these low temperatures. Compact and modular liquid $^3$He immersion cells specially designed 
for cryogen-free refrigerator are routinely used at the National High Magnetic Field Laboratory's (NHMFL or MagLab) 
High \textit{B}/\textit{T} Facility to ensure an improved thermal coupling of the on-chip electrons 
within nano-sized low dimensional devices to the mixing chamber temperatures in the magnetic fields 
up to 14~T.\cite{Zheng-1, Zheng-2, Zheng-3, MIT-1, MIT-2}

However, achieving temperatures \textit{below} 10 mK on cryogen-free systems is difficult as the 
cooling power relies on a pulse tube which inevitably introduces mechanical vibrations at 
low frequencies that are challenging to isolate from the experiment. These vibrations, when 
combined with the magnetic field required for the demagnetization technique to achieve 
temperatures of about 1 mK,\cite{Pobell} cause significant eddy current heating and thus 
pose a significant experimental challenge for employing demagnetization in cryogen-free 
systems when compared to traditional wet refrigerators. Despite these challenges, 
there are various types of nuclear demagnetization methods that have been implemented 
so far in cryogen-free systems in different scientific groups around the world.  

\textit{Parallel nuclear demagnetization of external electrical leads.} \cite{Basel2010, Basel2017}
Cooling to temperatures below 10 mK is achieved by employing a series of demagnetization stages, 
where each measuring/control lead attached to a sample is routed through its own, individual 
nuclear refrigerator. This arrangement removes the less efficient cooling path through an 
electrical insulator and replaces it with electronic Wiedemann-Franz cooling. However, 
the center of the final magnetic field is outside of the sample region, thus limiting the 
study of the sample to low magnetic fields unless a separate magnet is used for the experiment.

\textit{On-chip nuclear demagnetization.} \cite{Basel2017-APL, Delft2019} This method 
employs a minimal amount of refrigerant that is incorporated into a nanoelectronic device. 
The refrigerant is connected electrically to the conduction electrons of the device, 
effectively bypassing the limitations of weak electron-phonon coupling and the inefficient 
thermal connection to off-chip wiring, particularly in high-impedance devices. 

\textit{Nuclear demagnetization of an off-chip bundle.}\cite{Helsinki2014, OneHundred2021, London2022, Grenoble2024} 
In this approach the cooling below 10 mK and the thermalization of the device are treated 
as distinct processes, similar to traditional wet cryogenic  systems where the nuclear 
refrigerator and the magnet are separated from the experiment. Consequently, a tortuous 
electronic conduction path connects the refrigerant nuclei to the sample, posing challenges 
for achieving efficient thermalization of the sample down to the lowest temperatures. 
This challenge is particularly pronounced in the case of nanodevices, where thermal 
link between the stage and the electrons within the device becomes notably weak.

Herein, a different approach for a nuclear demagnetization experiment in a cryogen-free 
refrigerator is described, where powder/flake copper is suspended in the pure liquid $^3$He 
 used as the thermal media in an immersion cell. 
The combination of both the magnetic refrigerant and the pure liquid $^3$He, for 
thermalization of the sample\cite{Schuhl1987} and connecting leads\cite{Pan1999,Xia2000}, facilitate the magneto-transport study of low-dimensional devices at 
ultra-low temperatures below 10~mK and in sizeable magnetic fields, 
$1\,\mathrm{T}\,\lesssim B\,\lesssim\,5\,\mathrm{T}$, corresponding to the 
final field after the demagnetization process. This approach was previously implemented in 
wet dilution refrigerators to study quantum fluids, \cite{lanc82, lanc84} but to date, has not 
been tested or deployed on cryogen-free platforms. The use of powder eliminates the conduction 
path along a copper bundle or wire, providing efficient thermal transfer directly to 
liquid $^3$He through the large total surface area. The electrical conduction between 
the copper flakes is sufficiently low due to a stearate coating layer, \cite{lanc84}  
hereby minimizing eddy current heating due to vibrations in the magnetic field. 
More specifically, the copper powder has much stronger internal magnetic field of about 
350~mT, Ref.~\onlinecite{lanc84}, compared to only 0.4 mT in the case of 
bulk copper. \cite{Pobell}
This field provides a significant 
heat capacity to the copper nuclear spin system even at a zero external magnetic field, 
thereby protecting the cell from parasitic heat load and extending the hold times after demagnetization process. 

The following sections provide the details of the design, assembly of instrumentation, and the outcomes of 
testing experimental cells which allow the response of the sample to be monitored while both the 
temperature and magnetic field are slowly varied.  
Specifically, a pure liquid $^3$He immersion cell with copper powder is constructed and tested 
in a Bluefors LD250 dilution refrigerator equipped with the 14~T superconducting magnet made by American Magnetics Inc. 
The data are evidence for achieving temperatures of $2\,\mathrm{mK}\,\lesssim\, T \,\lesssim\, 5\,\mathrm{mK}$ in liquid $^3$He in the magnetic fields of $B\,\sim 1\, \mathrm{T}$ for a time $t \,\sim\, 10\,\mathrm{h}$.


\section{\label{sec:Cell}Copper powder demagnetization cell in a cryogen-free cryostat\protect}

\begin{figure}[t]
    \centering
    \includegraphics[width=\columnwidth]{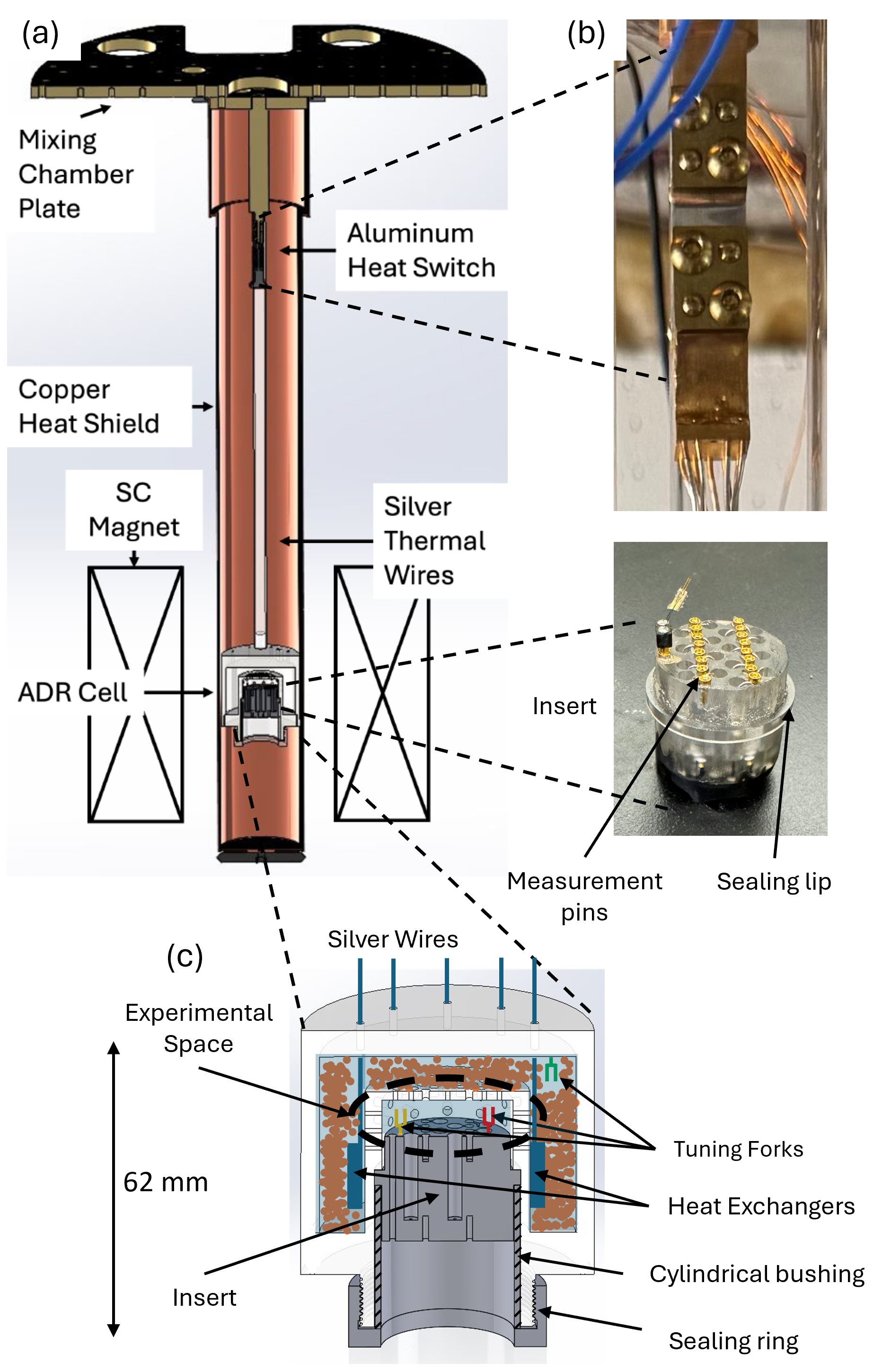}
    \caption{\label{fig:layout} (a) Layout of the cell and cryostat components. 
(b, \textit{upper photo}) The aluminum heat switch is thermally anchored to gold-plated copper at the top 
and gold-plated silver at the bottom using brass screws and Mo washers. (b, \textit{lower photo}) The insert with connected tuning fork used for thermometry. A sample/device can be mounted on the insert with a standard 16-pin DIP socket. The pockets near each pin accommodate silver sinter heat exchangers for an efficient thermalization. 
(c) Sketch of the cell incorporating copper powder in the ``demagnetization'' part. 
The experimental space is located in the center to provide additional thermal shielding.
Three tuning fork thermometers are located in different parts of the cell to control 
the liquid He-3 level and its temperature.
}
\end{figure}

The configuration of the low temperature stage is shown in Fig.~\ref{fig:layout}(a). The cell, whose main outer components are made with standard polycarbonate, is entirely positioned in the bore of a superconducting magnet so the sample is in the maximum field region. The cell is 
held by the polycarbonate tubes (9.5 mm OD and 6.3 mm ID, not shown in Fig.~\ref{fig:layout}) that are attached to the mixing chamber. 
Several annealed silver wires provide the thermal link to the mixing chamber 
through the aluminum heat switch shown in Fig.~\ref{fig:layout}(b,~\textit{upper photo}). 
The switch consists of four aluminum sheets thermally anchored to the copper leafs using brass 
screws with molybdenum washers. The contribution of the thermal resistance of 
the aluminum oxide layer to pre-cooling times in high magnetic fields is 
discussed in the \textit{Supplementary Material}. 
The heat switch is located 
in the fringe field region so it switches during the ramping 
of the magnetic field at a center strength of about 4~T. The fringe field at 
the position of the heat switch is above 10~mT in this case, therefore, it is
high enough to destroy the superconducting state of aluminum. 
The home-made heat shield is attached to the mixing chamber plate to protect 
the parts of the setup below the heat switch from radiation heat load from higher 
temperature parts of the cryostat. Importantly, no construction changes are made 
to the cryostat.

The cell design is shown in Fig.~\ref{fig:layout}(c). It contains two main parts, 
the ``demagnetization space'' and the ``experimental space''. 
The ``demagnetization space'' is filled with copper powder. 
It is initially compressed to about 30\% packing fraction using a hydraulic press. 
The resulting blocks are then broken into smaller pieces and added to the cell. 
An array of annealed silver wires with silver sinter heat exchangers are placed 
symmetrically into the copper powder for initial pre-cooling of the copper powder 
and liquid $^3$He before the demagnetization process. The ``experimental space'' is 
positioned at the center of the magnetic field region. The insert, shown in Fig.~\ref{fig:layout}(c) as the gray part in the middle and also in Fig.~\ref{fig:layout}(b,~{\textit{lower photo}}), is designed to hold a sample/device in the ``experimental space'' using electrical pins secured to the insert with silver epoxy to make it compatible with a standard 16-pin DIP socket. A light layer of Cello-Seal\textregistered{} C601-100 vacuum grease is applied to the top edge of the sealing lip, and then the insert is positioned into the cell body by a polycarbonate cylindrical bushing which is held in place by a polycarbonate threaded ring using light finger-only torque. Each measurement wire is equipped with its own silver sinter contact housed in a pocket on the insert, allowing for rapid thermalization of the electrons to the liquid $^3$He bath. During this experiment the two tuning fork (TF) resonators are used in the 
``experimental space'' to measure the temperature of the liquid $^3$He.
Importantly, tuning fork thermometry does not depend on the applied 
magnetic field up to at least 14 T,\cite{Woods2023}  so the temperature 
measurements at different magnetic fields do not require re-calibration of the thermometers.

In total, the cell contains about 66 g of copper powder with a surface area of 
about 61~m$^2$. The surface area of the silver sinters inside the cell is about 
13~m$^2$. The amount of $^3$He required to fill the cell is about 16~liters 
of gas at room temperature. It is important to determine the total 
entropy of the nuclear spin system and liquid $^3$He in the cell at the 
conditions prior to demagnetization. The entropy of non-interacting copper 
nuclear spins is given by\cite{Pobell}
\begin{equation} \label{eqn:cu-entropy-1}
\frac{S_{\mathrm{Cu}}}{Nk_B} = \ln{\left(\frac{\sinh (2x)}{\sinh (x/2)}\right)} + (x/2)\left[\coth(x/2) - 4\coth (2x)\right]
\end{equation}
\noindent
where $x = 0.54~[\mathrm{mK}~\mathrm{T^{-1}}]\left(\frac{B}{T}\right)$ in the case of copper.
The entropy of liquid $^3$He is given by\cite{lanc82}
\begin{equation} \label{eqn:cu-entropy-2}S_{\mathrm{3He}} = n\cdot 24~[\mathrm{J}~\mathrm{K^{-2}}~\mathrm{mol^{-1}}]\cdot T\;\;\;.
\end{equation}

\begin{figure}[t]
\centering
\includegraphics[width=\columnwidth]{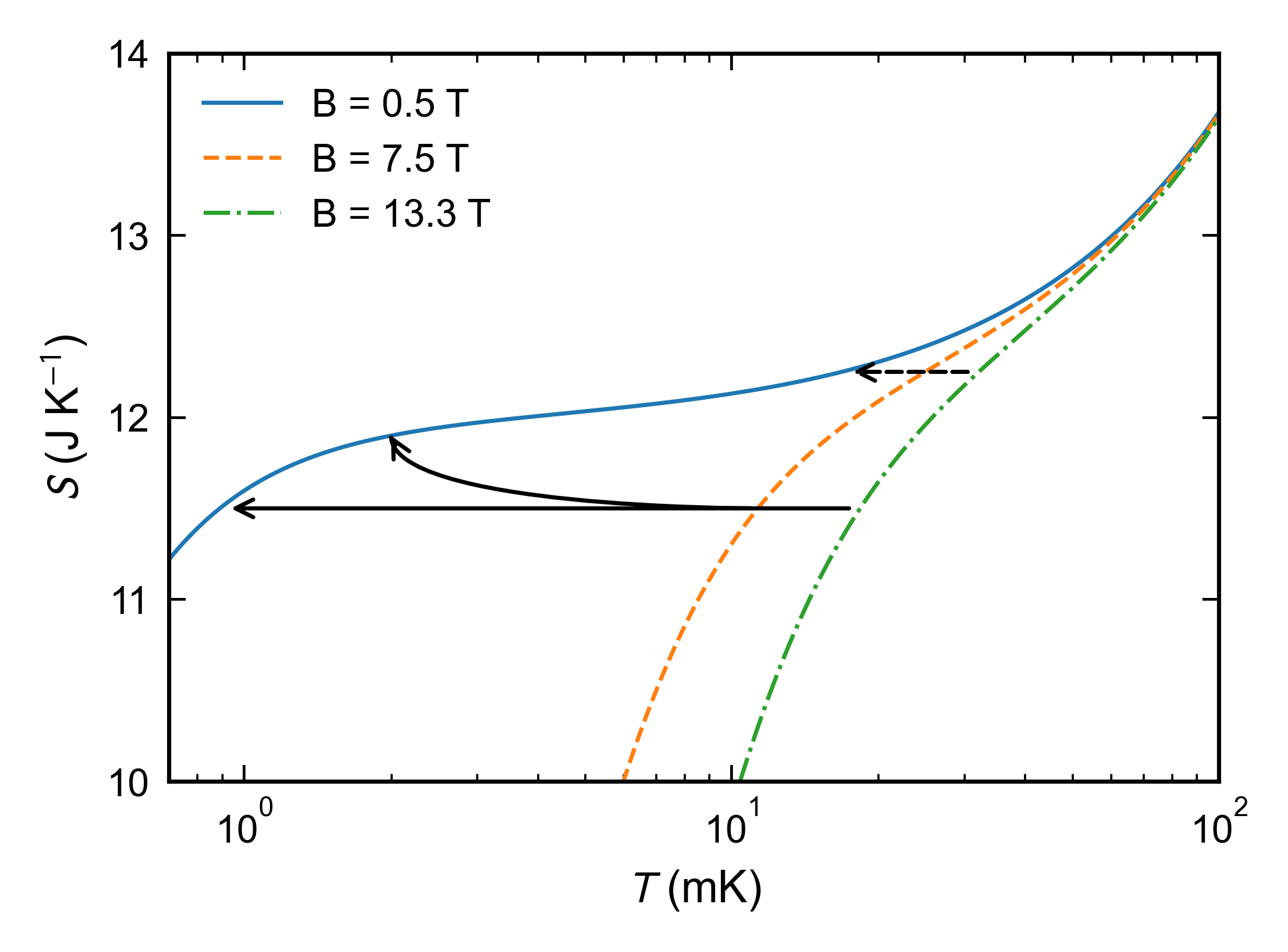} 
\caption{\label{fig:entropy} Total entropy of the 
``liquid $^3$He + copper powder'' system calculated at different magnetic fields 
using cell parameters. As shown by solid horizontal arrow, the demagnetization process is 
efficient when the cell is precooled below about 20~mK, if the applied field is 13.3~T. The curved solid arrow shows the demagnetization process when the system experiences an additional heat load $Q \,\sim\, 10\,\mathrm{nW}$.
The dashed arrow shows an inefficient demagnetization process when the starting 
temperature is too high. At smaller magnetic field the lower starting 
temperatures are required as shown by light orange line. }
\end{figure}

The temperature dependence of the total entropy $S = S_{\mathrm{Cu}} + S_{\mathrm{3He}}$ is 
shown in Fig.~\ref{fig:entropy} in different magnetic fields. 
The arrows represent the demagnetization process from different 
starting conditions. It shows that the process would be efficient for the 
current cell configuration if initial temperatures below 20 mK could be 
achieved in 14~T magnetic field even if an additional heat load $Q \,\sim\, 10\,\mathrm{nW}$ is present.

\section{\label{sec:Thermometry}Thermometry\protect}

The process of using tuning forks immersed in pure liquid $^3$He to measure temperature 
involves tracking the linewidths of their resonances.\cite{Woods2023} 
The tuning fork dimensions are characterized by their length $L$, thickness $e$, 
and width $\ell$. The linewidth $\Delta$\textit{f} reflects the hydrodynamic forces 
acting on the tuning forks  based on the Euler-Bernoulli beam theory, which 
yields the expression\cite{ZhangSensors, Sader1998, Sader2005, Sader2010}

\begin{equation} 
    \label{eqn:delta omega = hydro}
    \Delta f_1 = f_0 \left(\frac{\pi \rho\ell}{4\rho_{q}e}\right)
    \frac{\Gamma_{\mathrm{Im}}(f_1)}{\left[1 + \left(\frac{\pi \rho\ell}{4\rho_{q}e}\right)\Gamma_{\mathrm{Re}}(f_1)\right]^{3/2}}\;\;\;,
\end{equation}

\begin{figure}[t]
    \centering
    \includegraphics[width=\columnwidth]{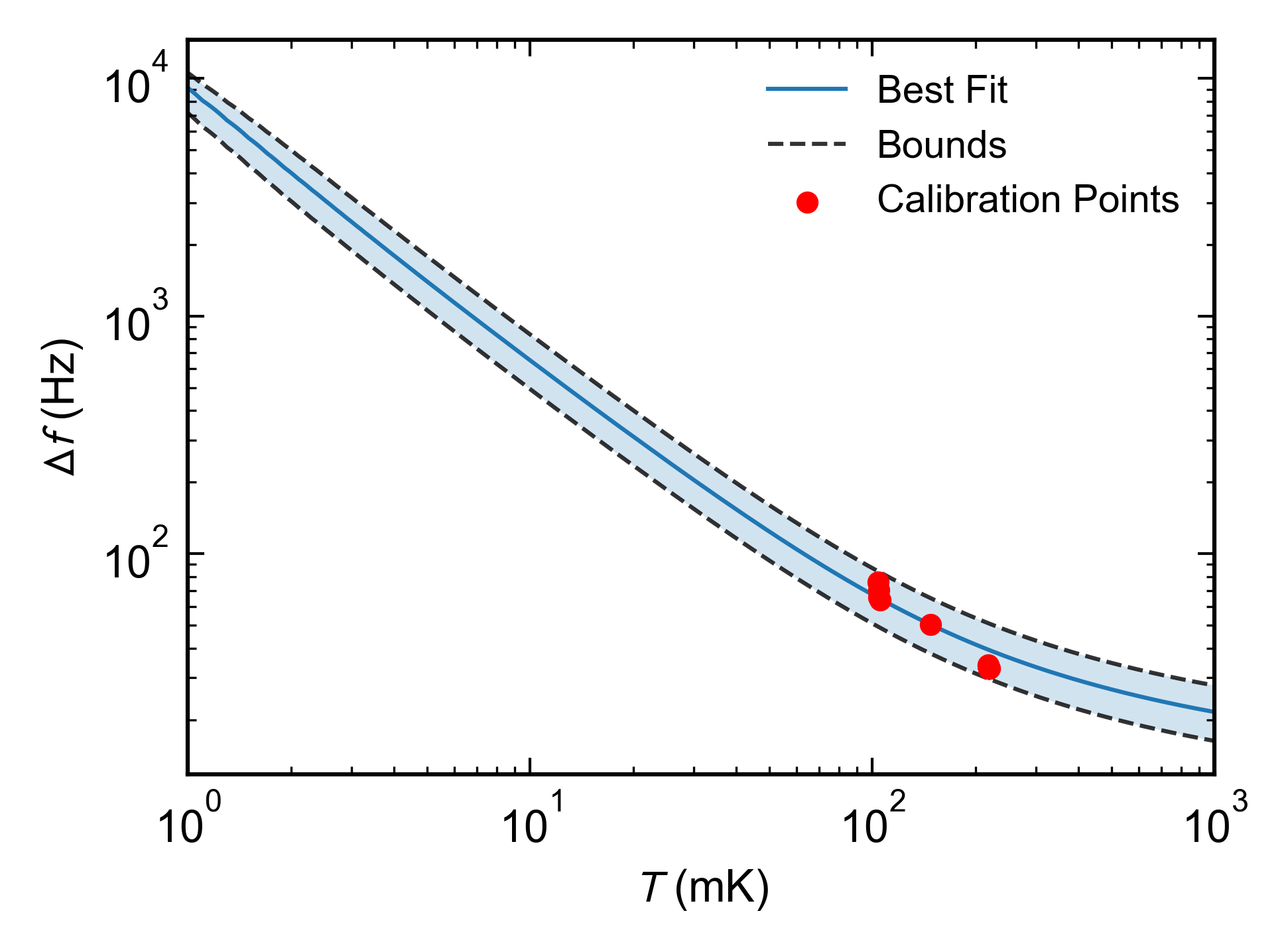}
    \caption{Red points represent the measured linewidths of the tuning forks  
    at known temperatures. The curve is the best fit from the hydrodynamic 
    model while the shaded region shows the uncertainty area, as described in the text. }
    \label{fig:calibration}
\end{figure}

\noindent
where $f_0 = \frac{\alpha_{1}^{2} \ell}{4 \pi L^{2}}\sqrt{\frac{E}{3\rho_{q}}}$ is the 
vacuum frequency, and $\Delta f_{1}$ and $f_{1}$ are from the first resonance mode due to 
the hydrodynamic damping. The other constants are the densities of quartz ($\rho_q$) and 
$^3$He ($\rho$), the Young's modulus of quartz (E), and $\alpha_1=1.875$ which is the 
first positive root of $1+\cos(\alpha_n)\cosh(\alpha_n)=0$.
In addition, $f_0$ and $f_1$ are related by the equation

\begin{equation} 
    f_{0}^{2} = f_{1}^{2} \left( 1 + \frac{\pi \ell}{4 \rho_{q} e}~\rho ~\Gamma_{\mathrm{hydro}}^{\mathrm{Re}}(f_{1}) \right)\;\;\;.
    \label{hydro omega_1}
\end{equation}

Finally, the dimensionless hydrodynamic function is given by
\begin{equation} \label{hydro func}
    \Gamma_{\mathrm{hydro}}(f) = \Gamma_{\mathrm{Re}} - i\Gamma_{\mathrm{Im}} = 1 + \frac{4 K_1(\sqrt{iRe})}{\sqrt{iRe} K_0(\sqrt{iRe})}\;\;\;,
\end{equation}
\noindent
where $K_1$ and $K_0$ are first-order modified Bessel functions of the second kind in 
addition to the Reynolds number $Re = \frac{\pi \rho f \ell^2}{2 \eta}$ where $\eta$ 
is the dynamic viscosity.

This equation provides the viscosity of the liquid $^3$He in which the tuning forks are 
immersed. Since the temperature dependence of liquid $^3$He viscosity is well established, 
these functions are used to convert the measured viscosities to the temperature of the liquid in its 
monotonic region from 1 K down to 1 mK.\cite{HUANG2012538}

The tuning fork response is calibrated above 100~mK, using the resistive thermometer 
anchored to the silver wires at the top of the cell (Fig.~\ref{fig:calibration}). 
The red points are taken after setting the mixing chamber temperature to different 
values and allowing the cell to thermalize through the heat switch in the superconducting 
state over the course of 24 hours. The thermometry model is then fitted against these 
calibration points by adjusting the dimension of the tuning forks within a physically 
plausible range to provide the best fit and thereby establish the bounds of the 
calibration points. This adjustment is shown in Fig.~\ref{fig:calibration} as 
a shaded region, which essentially defines the uncertainty of the temperatures. 

Since the resistive thermometer used for the calibration process loses sensitivity 
below 100~mK, the fits to the model are performed above this temperature, and then 
the hydrodynamic model is used to capture the temperature dependence of 
the linewidths down to about 1 mK.\cite{Woods2023} 


\section{\label{sec:Performance}Performance\protect}

\begin{figure} [t]
    \centering
    \includegraphics[width=\columnwidth]{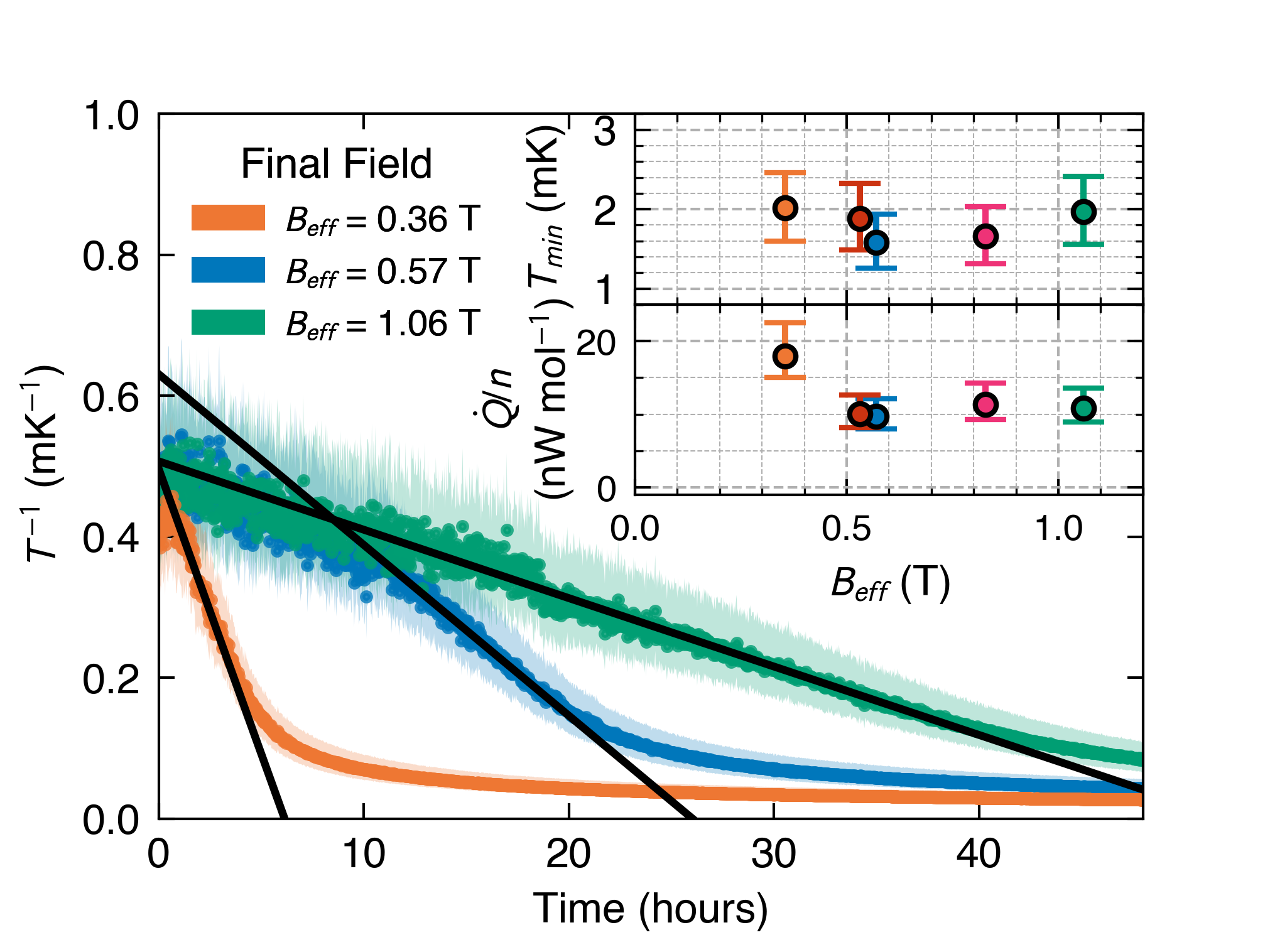}
    \caption{Inverse temperature as a function of time at different magnetic fields. 
    The slopes of the linear fits indicate a heat load of a few nW/mol. 
    \textit{Inset}: Minimum temperature and heat load per mole at different magnetic fields.}
    \label{fig:1_over_T}
\end{figure}

Multiple cooling cycles have been performed in three processes: magnetized precooling, 
demagnetization, and holding time. After applying a magnetic field of 13.3~T, 
the precooling process takes about 4~days for the cell to reach 16~mK. 
At this point the demagnetization process is expected to be efficient (Fig.~\ref{fig:entropy}).
The magnetic field is then ramped down at 0.2~mT/s from 13.3~T to 5~T, and at 0.1~mT/s 
from 5~T down to final fields. This process is quasi-adiabatic and shows continuous cooling down to 2~mK. The \textit{B}/\textit{T} ratio is nearly constant down to at least 5~mK (see \textit{Supplementary Material}).
Following the end of the demagnetization, the cell warms from its base temperature 
over the course of 1-2 days depending on the final demagnetization field. The time traces of the inverse temperature are shown in Fig.~\ref{fig:1_over_T}.

Under a constant, post-demagnetization heat load $\dot{Q}$, the temperature $T$ of the 
nuclear spins is to evolve with time as given by \cite{Pobell} 

\begin{equation} \label{1_over_T} 
T^{-1} = T^{-1}_{\mathrm{min}} - \frac{\mu_{0}\,\dot{Q}}{n\,\lambda_{n}\, B_{\mathrm{eff}}^{2}}\;t
\end{equation}
\noindent
where $\mu_0$ and \textit{n} are the vacuum permeability and the amount of copper nuclei (in moles), respectively.
The quantity $\lambda_{n}/\mu_{0}=5N_{0}\mu_{n}^{2}g_{n}^2/4k_{B}$ is reported to be 
3.22 $\upmu$J K T$^{-2}$ mol$^{-1}$ for copper nuclei.\cite{Pobell} 
The internal magnetic field of copper is higher than the value reported in bulk metal and 
has been measured to be  $\approx350$~mT, which provides the effective magnetic field 
$B_{\mathrm{eff}}^{2} = B^2+B^2_{\mathrm{int}}$. \cite{lanc84, Pobell}

\begin{figure} [t]
    \centering
    \includegraphics[width=\columnwidth]{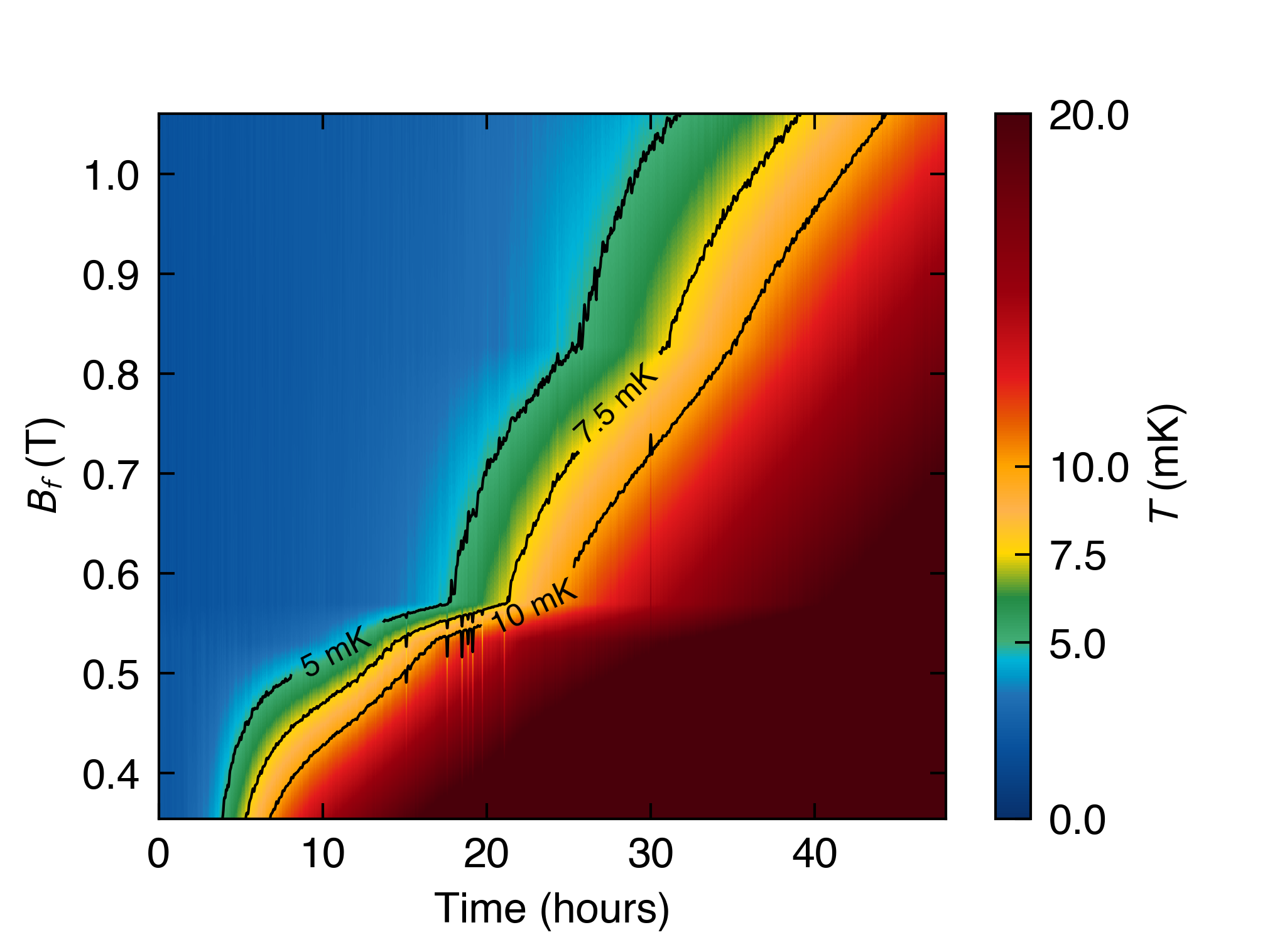}
    \caption{A final magnetic field vs time contour plot of the temperature showing the performance/hold time during the post demagnetization warm-up. The black lines show the equal temperature lines across the various runs.}
    \label{fig:T_colormap_post_demag}
\end{figure}

The measured $1/T$ values have a linear time dependence (Fig.~\ref{fig:1_over_T}, solid lines). 
Assuming that the liquid $^3$He is well coupled to the electron temperature of the copper due 
to a high contact area, the minimum achieved temperature and the average heat load on the cell 
are calculated using Eq.~(\ref{1_over_T}) as shown in the inset of Fig.~\ref{fig:1_over_T}. 
This analysis suggests our cell has reached a minimum temperature of $\lesssim2$~mK. 
Due to uncertainty in the calibration of the tuning fork thermometry, 
the uncertainties are $\pm$ 0.5 mK.
Two demagnetization runs with final fields $B_f=60$~mT and $B_f=400$~mT 
have higher minimum temperatures than the adiabatic, constant $B/T$ ratio estimate, 
and these outcomes are associated with the presence of higher heat loads near the 
end of the demagnetization process due to flux jumps in the main magnet at 
low current levels.\cite{magnet}

A noteworthy outcome is the heat load following the end of each demagnetization 
(Fig.~\ref{fig:1_over_T}, \textit{inset}) does not change significantly with 
the final applied field, indicating negligible eddy current heating.
This result directly reflects a key design objective of the cell and therefore strongly 
motivates further refinement and iteration to achieve improved performance.

Lastly, there is a clear deviation from expected linear dependence when 
$1/T \lesssim 0.1$~ mK$^{-1}$, Fig.~\ref{fig:1_over_T}. One of the possibilities might be 
that the plastic cell body is not being cooled efficiently and remains at temperatures higher than the 
mixing chamber, thereby leading to an additional radiation heat load. 
Presently, the source of this behavior is being investigated. 

In Fig.~\ref{fig:T_colormap_post_demag}, a color plot of the temperature time evolution after demagnetization to various final magnetic fields is shown. Voltage spikes in the magnet at low currents/magnetic fields \cite{magnet} has caused shorter than expected hold times at final field below 500~mT. The holding time below 5~mK and above 500~mT ranges from 18~hours to 30 hours. 


\section{\label{sec:Conclusion}Conclusion\protect}

``In-cell'' magnetic cooling below 2~mK has been successfully demonstrated using 
a cryogen‑free refrigerator. Remarkably, no eddy‑current heating is observed, 
a result of particular importance for pulse‑tube–cooled systems. 
The prototype cell required only 66~g (1.04~mol) of copper, representing a substantial 
reduction relative to traditional bundle‑stage designs that typically employ several 
kilograms of material. Crucially, this compact approach requires no modifications 
to the existing cryostat and thus provides a practical route for cooling nanoelectronic 
materials and devices to sub‑5~mK temperatures in finite magnetic fields. 
As a result, the accessible parameter space for ultralow temperature quantum experiments 
is significantly expanded, enabling broader adoption by the research community.

\begin{acknowledgments}
We gratefully acknowledge Bill Malphurs and UF Physics Machine shop for producing the 
custom parts used in the project, James Hamlin for providing the hydraulic press used for 
packing the copper powder, and Yoon Lee for enlightening discussions.

This work was performed at the National High Magnetic Field Laboratory High $B/T$ Facility 
on the campus of the University of Florida. The National High Magnetic Field Laboratory is 
supported by the National Science Foundation through Grants DMR-1644779 and DMR-2128556 
and the State of Florida. This work was partially supported by the National High Magnetic 
Field Laboratory through the NHMFL User Collaboration Grants Program (UCGP).
\end{acknowledgments}

\section*{Data Availability Statement}

The data that support the findings of
this study are openly available in
Zenodo at
https://doi.org/10.5281/zenodo.20286028.

\nocite{*}
\bibliography{referenceFile}

\end{document}